\documentclass[10pt, journal]{IEEEtran}

\usepackage{latexsym}
\usepackage{amscd}
\usepackage{amsfonts}
\usepackage{float}
\usepackage{longtable}
\usepackage{xspace}
\usepackage{stmaryrd}
\usepackage[utf8]{inputenc}
\usepackage{textcomp}

\DeclareMathAlphabet{\mathsl}{OT1}{cmr}{m}{sl}
\DeclareMathAlphabet{\mathsc}{OT1}{cmr}{m}{sc}
\DeclareMathAlphabet{\mathslbf}{OT1}{cmr}{bx}{sl}
\DeclareFontFamily{OT1}{pzc}{}
\DeclareFontShape{OT1}{pzc}{m}{it}%
             {<-> s * [1.150] pzcmi7t}{}
\DeclareMathAlphabet{\mathscript}{OT1}{pzc}{m}{it}

\newcommand{\secref}[1]{Section~\ref{#1}}

\newcommand{\authnote}[2]{
\ifnum\authnotes=1 
  \begin{center}
    \fbox{\begin{minipage}{5.7in}
      \textbf{#1 says:} #2
    \end{minipage}}
  \end{center} 
\fi
}

\newcommand{\vecx}{\mathbf{x}}
\newcommand{\vecy}{\mathbf{y}}

\newcommand{\pmul}{\ensuremath{\pi_{\mathsf{DM}}}\xspace}

\newcommand{\pmmul}{\ensuremath{\pi_{\mathsf{DMM}}}\xspace}

\newcommand{\shareq}[1]{\ensuremath{\llbracket{#1}\rrbracket_{_q}}\xspace}

\newcommand{\ptrunc}{\ensuremath{\Pi_{\mathsf{Trunc}}}\xspace}

\newcommand{\pmatinv}{\ensuremath{\Pi_{\mathsf{MatInv}}}\xspace}

\usepackage{graphicx}
\usepackage{amsmath}
\usepackage{amssymb}
\usepackage{tcolorbox}
\usepackage{color}
\usepackage{url,cite}
\usepackage{enumitem}
\usepackage{algpseudocode}

\begin{document}
\title{Protecting Privacy of Users in\\ Brain-Computer Interface Applications}

\author{Anisha Agarwal,
        Rafael Dowsley,
        Nicholas D.~McKinney,
        Dongrui Wu,\\
        Chin-Teng Lin,
        Martine De Cock,
        and~Anderson C. A. Nascimento
        
\thanks{\textcopyright ~2019 IEEE.  Personal use of this material is permitted.  Permission from IEEE must be obtained for all other uses, in any current or future media, including reprinting/republishing this material for advertising or promotional purposes, creating new collective works, for resale or redistribution to servers or lists, or reuse of any copyrighted component of this work in other works.}
\thanks{M. De Cock and A. Nascimento are with the School of Engineering and Technology, University of Washington Tacoma, WA 98402, USA, e-mail: \{mdecock,andclay\}@uw.edu.}%
\thanks{
R. Dowsley is with the Department of Computer Science, Bar-Ilan University, Israel. Email: rafael@dowsley.net. He is supported by the BIU Center for Research in Applied Cryptography and Cyber Security in conjunction with the Israel National Cyber Bureau in the Prime Minister's Office.
This work was mostly done while he was with the Department of Computer Science, Aarhus University. He has received funding from the European Research Council (ERC) under the European  Union's  Horizon  2020  research  and  innovation programme  under  grant  agreement  No  669255  (MPCPRO).}%
\thanks{A. Agarwal and N. McKinney were with the School of Engineering and Technology, University of Washington Tacoma, WA 98402, USA, e-mail: \{anisha3,mckinnnd\}@uw.edu.}%
\thanks{D. Wu is with the Key Laboratory of the Ministry of Education for Image Processing and Intelligent Control, School of Automation, Huazhong University of Science and Technology, Wuhan, China, email: drwu09@gmail.com.}%
\thanks{C.-T. Lin is with the School of Software, University of Technology, Sydney, Australia,
email: Chin-Teng.Lin@uts.edu.au.}%
\thanks{M.~De Cock is a guest professor at the Department of Applied Mathematics, Computer Science, and Statistics, Ghent University, 9000 Ghent, Belgium, e-mail: martine.decock@ugent.be.}
}

\markboth{IEEE Transactions on Neural Systems and Rehabilitation Engineering}%
{Shell \MakeLowercase{\textit{et al.}}: Bare Demo of IEEEtran.cls for IEEE Journals}


\maketitle

\begin{abstract}
Machine learning (ML) is revolutionizing research and industry. Many ML applications rely on the use of large amounts of personal data for training and inference. Among the most intimate exploited data sources is electroencephalogram (EEG) data, a kind of data that is so rich with information that application developers can easily gain knowledge beyond the professed scope from unprotected EEG signals, including passwords, ATM PINs, and other intimate data. The challenge we address is how to engage in meaningful ML with EEG data while protecting the privacy of users.

Hence, we propose cryptographic protocols based on  Secure Multiparty Computation (SMC) to perform linear regression over EEG signals from many users in a fully privacy-preserving (PP) fashion, i.e.~such that each individual's EEG signals are not revealed to anyone else. To illustrate the potential of our secure framework, we show how it allows estimating the drowsiness of drivers from their EEG signals as would be possible in the unencrypted case, and at a very reasonable computational cost. Our solution is the first application of commodity-based SMC to EEG data, as well as the largest documented experiment of secret sharing based SMC in general, namely with 15 players involved in all the computations.  
\end{abstract}


%
%

\section{Introduction}
The application potential of Brain-Computer Interfaces (BCIs) is vast, going far beyond medicine and research into areas such as education, gaming, entertainment, wellness, and personalized marketing. 
The emergence of consumer-grade, low-cost BCIs and corresponding software development kits\footnote{E.g.~\url{https://www.emotiv.com/}, \url{http://neurosky.com/}, \url{https://myndplay.com/}} is bringing the use of BCI within reach of application developers. They can capture neural signals, extract features from them, and subsequently use these extracted features to train and use machine learning (ML) models for all kinds of prediction and inference tasks. These include inferring emotions, sexual preferences and religious beliefs of individuals, detecting preferences of customers, measuring concentration, or estimating levels of drowsiness in drivers of cars \cite{bonaci2015susceptible,inzlicht2009neural,poel2010guessing,doborjeh2018attentional,qian2018brain,wu2017driver}. 

While many BCI-applications can be, and are, developed with a benign intent of enriching and improving the quality of human life, giving access to a user's brain signals, or features extracted from them, can seriously harm the user's privacy. Brain spyware has for instance been used to infer users' 4-digit PINs, bank information, month of birth, location of residence, and whether they recognized a presented set of faces \cite{martinovic2012feasibility}. The impact of brain malware that can infer very intimate information about users, such as emotions, prejudices, religious and political beliefs, etc.~and subsequently use that information to manipulate users, could be severe \cite{bonaci2014app}.

The awareness of the need for protecting the privacy of individuals and their data in ML applications has increased substantially over the last few years, as witnessed for instance in 
the National Privacy Research Strategy put forward by the National Science and Technology Council (Jun 2016)\footnote{\url{https://www.nitrd.gov/PUBS/NationalPrivacyResearchStrategy.pdf}}, 
the recommendations of the Commission on Evidence-Based Policy Making (Sep 2017)\footnote{\url{https://www.cep.gov/content/dam/cep/report/cep-final-report.pdf}}, 
and ACM's statement on preserving personal privacy (Mar 2018) \cite{ACMPrivacy:2018}.
Sensitive data includes user generated content on social media, patient healthcare records, genetic information, and -- without a doubt -- neural information such as recorded by EEG signals. There is plenty of evidence that anonymizing data does not offer sufficient protection \cite{sweeney2007systems}. In this paper we therefore focus on the use of cryptography, in particular Secure Multiparty Computation (SMC) \cite{CDN2015}, to ensure, in a mathematically provable way, that the EEG data of individuals used in ML applications is not revealed to anyone but themselves, while still being able to do meaningful computations over that data.

To this end, we propose cryptographic protocols for fully privacy-preserving linear regression (PPLR) with data from EEG signals, and their implementation in Lynx \cite{Lynx}, a framework for SMC based on additive secret sharing. Our methods are applicable in any application that requires training an LR model from EEG data. In this paper, 
we demonstrate our protocols for estimating the drowsiness of drivers, which is the cause of 1000s of fatal crashes each year.\footnote{\url{https://www.cdc.gov/features/dsdrowsydriving/index.html}}
We consider two different scenarios. In the \textit{first scenario}, a set of \textit{source drivers} work together to train an LR model in a distributed fashion (many-party SMC). Throughout this process, none of the drivers can see the data from the other drivers in an unencrypted way at any point. At the end of the protocol, all source drivers hold encrypted shares of the trained model, and a \textit{target driver} can obtain a prediction for his data by engaging in a cryptographic protocol with all of the source drivers (many-party SMC). In the \textit{second scenario}, the target driver has calibration data that can be leveraged to train a personalized and more accurate model. The target driver engages in a separate cryptographic protocol with each of the source drivers (2-party SMC) to train LR models, namely as many models as there are source drivers. Each model is trained on data from one source driver, as well as on some of the calibration data from the target driver. 
As before, any individual's EEG data is not disclosed to anyone else. Furthermore, at the end of the training protocol, no single user knows any of the trained models. Instead, knowledge about the models is split into encrypted shares that are ``owned'' in a distributed fashion by the drivers. Finally, the target driver can obtain a prediction for his data, as an average of the predictions by all trained models, through engaging in a cryptographic protocol with all the source drivers.

Our secure framework allows to estimate the drowsiness of drivers as would be possible in the unencrypted case, and scales well with the number of drivers. It is the first application of commodity-based SMC to EEG data, as well as the largest documented experiment of secret sharing based SMC in general, with 15 players involved in all the computations.  

In this paper, we focus on privacy-preserving machine learning (PPML) techniques based on SMC. There are  alternative paradigms for obtaining such a goal. They differ -- among other things -- in how much information is leaked during training and deployment of the ML models.  
In this paper, we work in the most restrictive of these scenarios -- where no leakage is allowed. We do not cover other approaches for PPML that leak some information, such as: differential privacy \cite{abadi2016deep}; trusted enclaves\cite{chandra2017securing}; federated learning \cite{bonawitz2017practical,shokri2015privacy}. To the best of our knowledge, none of these approaches have previously been applied to training ML models based on EEG data either.

%
%

\section{Related work}
BCI technology is gradually becoming more ubiquitous. On the academic side, great progress was made in the development of technology for ``mind reading'' from fMRI activation patterns. Among others, this includes recent works by Wang et al.~\cite{wang2017predicting} who successfully trained a ridge regression model to identify complex thoughts, such as, ``The witness shouted during the trial'', and by Du et al.~\cite{du2017sharing} who presented a method for identifying what a user is looking at, just by monitoring their brain activity. At the same time, a variety of neural engineering companies have already introduced inexpensive, consumer-grade BCI devices for measuring brain activity in the form of EEG signals, as well as so-called BCI App Stores to facilitate adoption of the BCI headsets \cite{bonaci2014app}, and efforts are underway to make the more informative magnetoencephalography (MEG) brain scanners wearable in practice \cite{boto2018moving}.

The access that BCI applications have to neural signals rightly raises privacy concerns. A well known threat are subliminal attacks in which users are exposed to visual stimuli for a duration that is too short for cognitive perception yet long enough to learn private information about the users based on their neural reactions to the visual stimuli (e.g.~brand logos) \cite{bonaci2015susceptible}. The data obtained in this way is valuable for example for phishing campaigns or ads.
Neural signals have also been used to elicit information about a person’s sexual orientation \cite{poel2010guessing} or religious beliefs \cite{inzlicht2009neural}. It is understood in the data science community that anonymization, i.e.~removing personally identifiable information from data before release, is not sufficient to protect the privacy of individuals, since it still leaves the data vulnerable to linkage attacks \cite{sweeney2007systems}. True protection can come from cryptographic techniques that allow computations over encrypted data, such as Fully Homomorphic Encryption (FHE) or Secure Multiparty Computation (SMC) \cite{CDN2015}.

Multiple approaches for secure LR have been proposed in the literature. Some are not based on SMC \cite{Chen2014,HalFieNar11,aono2015fast, nikolaenko2013privacy, karr2009privacy, sanil2004privacy}, and some use SMC like we do \cite{karr2005secure, du2001privacy,AISec:CDNN15,Gascon2017, DuCheHan04}. Several existing approaches assume that the data is vertically partitioned \cite{karr2009privacy, sanil2004privacy, Gascon2017}, hence it can not be used for the application that we study in this paper, in which each user has the information about his own EEG signals (i.e., horizontally partitioned data).

\textit{Homomorphic encryption (HE) based approaches.}
Hall et al.~\cite{HalFieNar11} achieve security in a two-party LR scenario, using HE on datasets over a finite field. The truncation protocol used in \cite{HalFieNar11} to scale down the finite field has a small problem which is documented in \cite{AISec:CDNN15}. The HE based method of Aono et al.~\cite{aono2015fast} outsources the computations to a server. The entire LR model is present at the server, and the client evaluates its data securely. Our approach differs from the above in various ways. Our method enables training and inference in a fully distributed fashion, i.e.~such that the coefficients of the trained LR model never have to be brought together in one place. Furthermore, our method allows an arbitrary number of parties, and computations are fast (less than 6 min for training an LR model with over 16,000 training examples distributed over 14 parties, and seconds for inference). Nikolaenko proposed a hybrid model for secure LR using HE and garbled circuits \cite{nikolaenko2013privacy}. While their approach does handle multiple parties, they upload encrypted data to a third party 
responsible for evaluating the model with the help of a semi-honest Crypto Service Provider. We eliminate the need of a third party to actively participate in the protocol while achieving better runtimes. 

\textit{SMC based approaches.}
Du et al.~\cite{du2001privacy} proposed an early approach for SMC based simple LR, i.e.~when there is only a single scalar predictor variable. The method we propose in this paper works for multivariable LR, which is far more common in practice.
Karr et al.~\cite{karr2005secure} provided a sketch for secure LR on horizontally partitioned data. They did not address important challenges that would need to be solved when implementing it in practice, such as how to perform matrix inversion in a secure manner and how to handle datasets with real numbers. Finally, Du et al.'s secure two-party approach~\cite{DuCheHan04} is different from ours in goal: they explore the trade-off between security requirements and efficiency based on the assumption that a dishonest party might be able to learn some information about the other party’s private data.  We are the first to implement cryptographic protocols for performing secret sharing based LR in which any number of parties can participate. To this end, we extend the PPLR technique by De Cock et al.~\cite{AISec:CDNN15} to $m$ parties. While the general protocols for LR in \cite{AISec:CDNN15} are similar to ours, the version presented in \cite{AISec:CDNN15} was not implemented for several parties. The implementation in \cite{AISec:CDNN15} was a simulation where all the parties were running within the same machine and did not include the delay due to the network connecting all the parties. To the best of our knowledge, our work is the largest documented experiment of privacy-preserving machine learning in terms of the number of parties. Moreover, our work includes private scoring, which was not present in \cite{AISec:CDNN15}. This paper extends our prior work to a real application with a realistic deployment, code that is re-usable and publicly available. For other recent work on the use of SMC for PPML (other than LR) we refer to \cite{Clifton:2002,aggarwal2008general,de2014practical,fritchman2018} and references therein. 

The implementation of new SMC protocols is facilitated by libraries that provide a general framework with a built-in implementation of cryptographic protocols for basic operations -- such as multiplication and comparison -- that one can use to implement more complex protocols in a modular fashion. In this paper we use the SMC library Lynx \cite{Lynx}, which is based on additive secret sharing. Other existing frameworks for SMC are Sharemind \cite{Bogdanov:2008}, FairPlay \cite{Ben-David:2008}, and Chameleon \cite{2018arXiv180103239S}. Both Sharemind and Chameleon are secret sharing based frameworks (like Lynx) developed in C++ whereas Fairplay uses both secret sharing and garbled circuits. Sharemind uses a fixed modulus of $2^{32}$ and, as it stands, is limited to computations on integers for three parties. Chameleon, a two-party framework with protocols similar to that of Sharemind, supports computations on floating point numbers in addition to integers. Like Lynx, Chameleon uses a trusted third party to generate correlated randomness. 
Fairplay, relies on a custom function definition language to define the boolean circuits. The need to learn a custom language makes it less user friendly. Chameleon and Sharemind are limited to 2 and 3 parties respectively. Fairplay can handle more than 3 parties but doing so comes at a substantial computational cost. The Lynx library that we use in this paper (see Section \ref{sec:architecture}) allows participation of an arbitrary number of parties. Lynx is designed to scale well with an increasing number of parties, among other things due to the use of a bulletin board functionality that enables efficient communication among many parties who are simultaneously involved in computations. To the best of our knowledge, ours is the first documented application of secret sharing based SMC for ML with computations done by more than 3 players.

We illustrate the power of our solution by applying it to the problem of privately estimating driver's drowsiness based on EEG data.
The U.S.~Department of Transportation reports that drowsy driving, i.e.~driving while experiencing sleepiness or fatigue, claimed 846 lives in 2014.\footnote{\url{https://www.nhtsa.gov/risky-driving/drowsy-driving}} 
According to the Centers for Disease Control and Prevention, up to 6,000 fatal crashes each year may be caused by drowsy drivers.\footnote{\url{https://www.cdc.gov/features/dsdrowsydriving/index.html}} The company \textit{Panasonic} has announced the release of an in-car system for driver drowsiness detection, through a combination of a camera and sensors which constantly measure blinking features, facial expressions, heat loss from the body, and illuminance \cite{Sandle:2017}. 
Depending on the detected level of tiredness, either the temperature in the car is changed (for moderate drowsiness), or an alarm is sounded (for severe drowsiness). Wu et al.~\cite{wu2017driver} have successfully trained linear regression models for inferring the level of drowsiness of drivers from their EEG signals, both in a setting where a model trained with data from $m$ source drivers is used to infer the drowsiness of a target driver, as well as in transfer learning settings where calibration data from the target driver is leveraged to personalize the predictive models, leading to more accurate drowsiness estimates. In this paper we show how regression models like those from Wu et al.~\cite{wu2017driver} can be trained and used in a fully privacy-preserving (PP) way, without any loss of accuracy, and at a very reasonable computational cost. A high level sketch of our work appeared previously \cite{aagarwal2018a}.

%
%

\section{Preliminaries}\label{sec:prelim}
In this section we introduce the notation for LR that we will adhere to in the paper, and we recall preliminaries about performing secure computations with additive secret sharings.

Throughout this paper we use capital letters such as $X$ to denote matrices, bold face letters such as $\vecy$ to denote vectors, and regular letters such as $y$ to denote scalar values. Let $X$ be an $n \times k$ matrix and $\vecy$ a vector of length $n$ as follows:
\begin{equation}\label{XandY}
X = \left( 
\begin{array}{c}
{\bf x}_1 \\
{\bf x}_2 \\
\ldots\\
{\bf x}_n 
\end{array}
\right)
{\rm \ and\ }
{\bf y} =
\left( 
\begin{array}{c}
y_1 \\
y_2 \\
\ldots\\
y_n
\end{array}
\right)
\end{equation}
Performing LR with $X$ and $\vecy$ means finding a coefficient vector
$\boldsymbol{\beta} = (\beta_0 \ \ \beta_1 \ \ \ldots \ \ \beta_k)$   
that minimizes
\begin{equation}\label{EMRformula}
\frac{1}{n} \sum\limits_{i=1}^n ((\beta_1 ({\bf x}_i)_1 + \beta_2 ({\bf x}_i)_2 + \ldots + \beta_k ({\bf x}_i)_k + \beta_0) - y_i)^2
\end{equation}
In a supervised ML application, $X$ and $\vecy$ contain information about training examples, where ${\bf x}_i$ is the input feature vector for the $i$th example and $y_i$ is the associated output. The goal is to leverage these training examples to predict the unknown outcome for a previously unseen input as accurately as possible by learning a linear function defined by the coefficient vector $\boldsymbol{\beta}$. The coefficients that minimize the mean squared error over the training examples (\ref{EMRformula}) can be computed as 
\begin{equation}\label{EMRformulasolved}
\boldsymbol{\beta} = (X^T X)^{-1}X^T {\bf y}
\end{equation}

In the scenarios that we are interested in, the data needed to train the LR model is not owned by a single party but is instead distributed across multiple parties who are not willing to disclose it. In other words, each of the parties has some of the entries of the matrix $X$ and the vector ${\bf y}$, and the parties are unwilling or unable to send their entries to each other or to a trusted third party to perform LR over the combined data.

To efficiently train LR models over distributed data in a PP way, we work in the commodity-based model~\cite{beaver1998one}. In this approach, there 
is a setup assumption about the existence of a Trusted Initializer (TI) that pre-distributes correlated random numbers during an initialization phase (which can happen far before the ML models are trained, even before knowing the training data) to the parties participating in the protocol. The TI is not involved in any other part of the execution and does not learn any input from the parties. The main advantage of the commodity-based approach is that it enables very efficient solutions with unconditional security. It has been used in the context of PPML \cite{AISec:CDNN15,david2015efficient,IEEETDSC:CDHK+17,fritchman2018}, as well as in other applications \cite{r99,beaver1995precomputing,dowsley2010two,IEICE:DMOHIN11,ishai2013power,IJIS:TNDMIHO15,IEEEIFS:DDGM+16}.

Throughout this paper, we perform secure computations using additive secret sharings over a finite field $\mathbb{F}_q$.  A value (number) $x \in \mathbb{F}_q$ is secret shared between  parties $p_1,\ldots, p_m$ by picking $x_{p_1}, \ldots, x_{p_m}  \in \mathbb{F}_q$ uniformly at random subject to the constraint that $x = \sum_{i=1}^m x_{p_i} \mod{q}$, and then revealing $x_{i}$ to $p_i$. This secret sharing will be denoted by $\shareq{x}$. Notice that from the point of view of any \textit{proper subset} of parties, no information about $x$ is revealed by the combination of their shares. A secret shared value can be revealed to one of the parties by sending him the shares of \textit{all} the other parties.

Given secret sharings $\shareq{x}$, $\shareq{y}$ and a constant $c$, it is trivial for the parties to compute secret sharings corresponding to $z=x+y$, $z=x-y$, $z=cx$, or $z=x+c$ by performing addition, subtraction, etc,~locally on the shares; the parties do not even need to communicate with each other to this end. These operations are respectively denoted by $\shareq{z} \gets \shareq{x}+\shareq{y}$, 
$\shareq{z} \gets \shareq{x}-\shareq{y}$, $\shareq{z} \gets c\shareq{x}$, and $\shareq{z} \gets \shareq{x}+c$. In the commodity-based model there is also a well-known protocol $\pmul$ to multiply the values of two secret sharings \cite{beaver1991efficient}, which has been generalized to a secure distributed matrix multiplication protocol $\pmmul$. At the start of the protocol $\pmmul$, the parties have element-wise secret sharings $\shareq{U}$ and $\shareq{V}$ of the matrices $U$ and $V$. At the end of the protocol, the parties have a secret sharing $\shareq{UV}$ of the product matrix $U V$. For a detailed description and a proof of security of this protocol, we refer to \cite{AISec:CDNN15, Dowsley16}.

For computing the coefficient vector using (\ref{EMRformulasolved}), besides matrix multiplication, we also need to 
compute the inverse of a covariance matrix. To do this in a PP fashion, we use a secure matrix inversion protocol that is based on a generalization of the Newton-Raphson division method to matrices \cite{AISec:CDNN15}. At the start of the protocol, which we denote as $\pmatinv$, the parties have shares of a covariance matrix $A$, and at the end of the protocol, they have shares of the inverted matrix $A^{-1}$. For details about the protocol $\pmatinv$, we refer to \cite{AISec:CDNN15}.

The protocols described above, and their security proofs, assume all computations are done with numbers from the finite field $\mathbb{F}_q$. In real-life applications, such as the BCI application of estimating driver drowsiness that we consider later in this paper, the inputs are real numbers. We therefore need a way to approximate computations with real numbers by computations with numbers from $\mathbb{F}_q$. To this end, 
we adapt the method of Catrina and Saxena \cite{catrina2010secure} for fixed-point representation of the numbers in the same way as was described in \cite{AISec:CDNN15}. Similarly, when secure multiplications are performed, we use the slightly modified version $\ptrunc$ of the truncation protocol of Catrina and Saxena \cite{catrina2010secure} that was presented in \cite{AISec:CDNN15}. 
For the computation of the results in Section \ref{sec:results}, numbers are represented with a $f=64$ bit decimal precision and a $e=64$ bit integer precision. For $q$, i.e.~the dimension of the field, we use the first prime value larger than $2^{e + 2f + 1}$ to allow the truncation protocol to work correctly and not result in an overflow during intermediate computations.

%
%

\section{Cryptographic Protocols}\label{sec:protocols}
We present a solution for PP training and inference with LR models in two different scenarios that are very relevant in practice, and both involve $m$ source parties and a target party:
\begin{itemize}[leftmargin=*,noitemsep,topsep=3pt]
    \item \textbf{Target-independent LR.} In the target-independent LR scenario, one LR model is trained with data from $m$ source parties, and used to make predictions about a target party. No data from the target party is used during the training phase. This scenario corresponds to ``Baseline 1'' in \cite{wu2017driver}.  
    \item \textbf{Target-calibrated LR.} In the target-calibrated LR scenario, $m$ LR models are trained, each with data from one of the $m$ source parties combined with some calibration data from the target party. Inferences for the target party are subsequently made by an ensemble of the trained LR models. This scenario corresponds to ``DAMF'' in \cite{wu2017driver}.
\end{itemize}

Both approaches are valuable in practice, and even more sophisticated techniques to leverage calibration data exist \cite{wu2017driver}. Our goal in this paper is not to investigate which of these techniques can lead to the most accurate predictions. Instead, our aim is to show that the computations needed to train and use such regression models can be performed in a fully PP way, i.e.~so that none of the parties involved has to disclose its data to anyone else in an unencrypted way. From the PP perspective, the two scenarios outlined above pose quite different challenges and require different protocols, which we describe in more detail below.

\subsection{Training for target-independent LR}\label{sec:trti}
\begin{figure}
\begin{tcolorbox}[
    standard jigsaw,
    opacityback=0,  
]
\begin{center}
\includegraphics[width=3in]{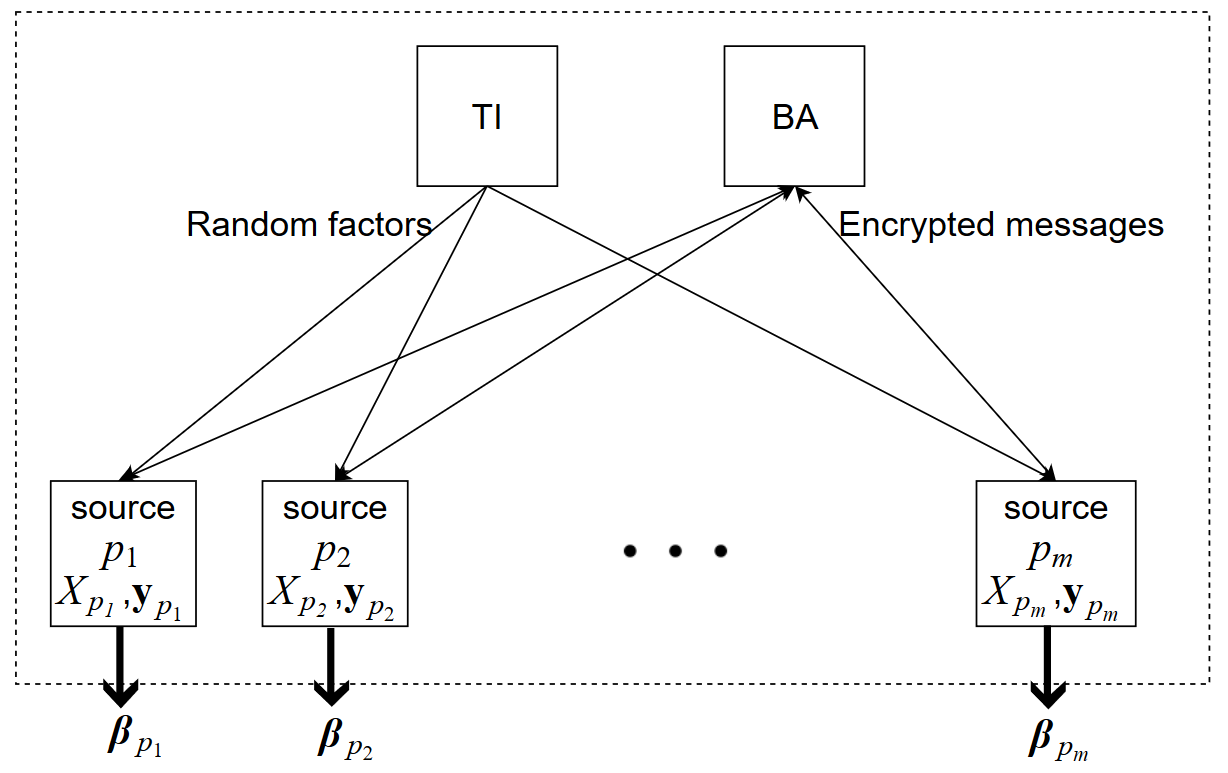}
\end{center}
\end{tcolorbox}
\caption{Training phase of privacy-preserving target-independent LR.}
\label{figure:SCEN1TRAIN}
\end{figure}

In the PP target-independent LR scenario, illustrated in Fig.~\ref{figure:SCEN1TRAIN}, each of the $m$ source parties $p_1, p_2, \ldots, p_m$ has its own rows of the matrix $X$ and corresponding entries of the vector $\vecy$ from (\ref{XandY}). Each party $p_i$ can construct its own $n_{p_i} \times k$ matrix $X_{p_i}$ and a vector $\vecy_{p_i}$ of length $n_{p_i}$, with $n_{p_i}$ the number of training examples held by party $p_i$ and $k$ the number of features. We take advantage of the fact that the data is horizontally partitioned in this way, and propose a protocol for PP training of a LR model with the data from all parties that is more efficient in this situation than the more general protocol from De Cock et al.~\cite{AISec:CDNN15}.
Our technique consists of the steps described below. 

\begin{figure}\begin{tcolorbox}[
    standard jigsaw,
    opacityback=0,  
]
\begin{center}
\includegraphics[width=3in]{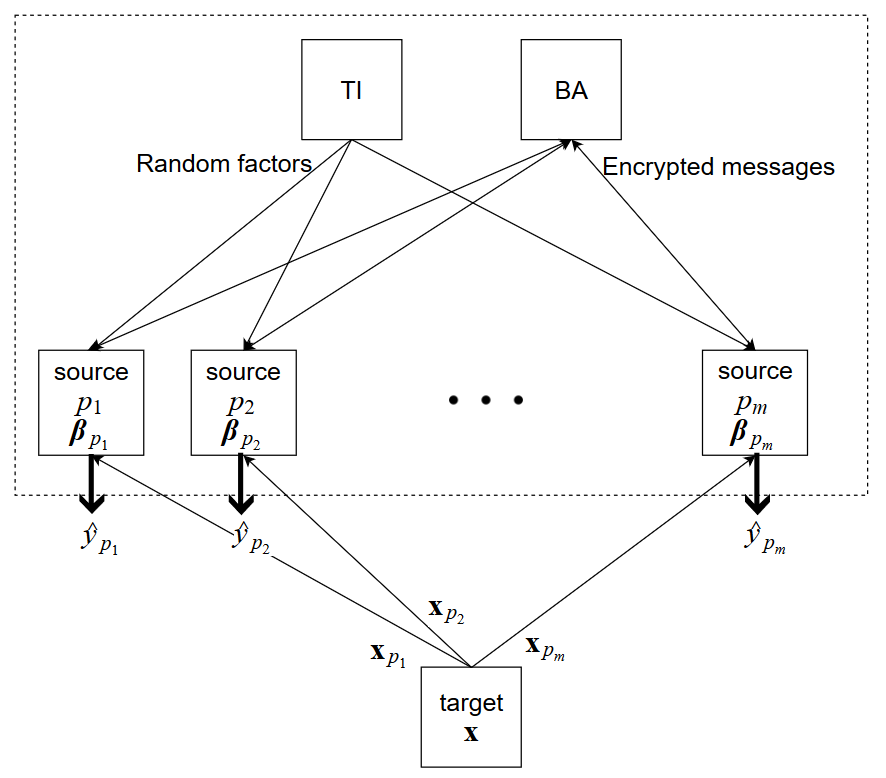}
\end{center}
\end{tcolorbox}
\caption{Inference phase of privacy-preserving target-independent LR.}
\label{figure:SCEN1TEST}
\end{figure}

\begin{enumerate}[leftmargin=*,noitemsep,topsep=3pt]
\item Offline phase: The TI distributes the necessary correlated random numbers to the parties. These random numbers will be needed for secure multiplications in the \ref{step4}th and \ref{step5}th step.

\item \label{step2} Local computation of $\shareq{X^TX}$: Each source party $p_i$ maps its fixed-point inputs to elements of a finite field $\mathbb{F}_q$ using the method described in \secref{sec:prelim}, and creates an $n_{p_i} \times k$ matrix $X_{p_i}$ and a vector ${\bf y}_{p_i}$ of length $n_{p_i}$, with $n_{p_i}$ the number of training examples owned by party $p_i$ and $k$ the number of features. Next, each source party \textbf{locally} computes the $k \times k$ matrix $(X_{p_i})^T (X_{p_i})$. It holds that
$
X^T X = (X_{p_1})^T (X_{p_1}) + \ldots + (X_{p_m})^T (X_{p_m}) \mod q.
$
In other words, each party $p_i$ now holds a secret share of the matrix $X^T X$ from Equation (\ref{EMRformulasolved}).

\item \label{step3} Local computation of $\shareq{X^T \vecy}$: each source party $p_i$ \textbf{locally} computes the $k \times 1$ matrix $(X_{p_i})^T (\vecy_{p_i})$. It holds that
$
X^T\vecy = (X_{p_1})^T (\vecy_{p_1}) + \ldots + (X_{p_m})^T (\vecy_{p_m}) \mod q.
$
In other words, each party $p_i$ now holds a secret share of the matrix $X^T\vecy$ from Equation (\ref{EMRformulasolved}).

\item \label{step4} Joint computation of $\shareq{(X^T X)^{-1}}$: the parties perform \textbf{joint} computations over their shares $\shareq{X^T X}$ from step \ref{step2} to compute shares $\shareq{(X^T X)^{-1}}$ of the inverse matrix $(X^T X)^{-1}$, using the protocol for covariance matrix inversion $\pmatinv$ mentioned in Section \ref{sec:prelim}, which in turn relies on the protocols for secure matrix multiplication $\pmmul$ and truncation $\ptrunc$.

\item \label{step5} Joint computation of $\shareq{\boldsymbol{\beta}}$: the parties perform \textbf{joint} computations over their shares $\shareq{(X^T X)^{-1}}$ from step \ref{step4} and $\shareq{X^T\vecy}$ from step \ref{step3} to compute 
shares $\shareq{\boldsymbol{\beta}}$ of the coefficient vector $\boldsymbol{\beta}=(X^T X)^{-1}X^T\vecy$. 
To this end, they perform distributed matrix multiplication between $(X^T X)^{-1}$ and $X^T\vecy$ by performing secure matrix multiplication $\pmmul$ and secure truncation protocol $\ptrunc$ operations.
In the end each party $p_i$ has a share $\boldsymbol{\beta}_{p_{i}}$ of the estimated regression coefficient vector
$\boldsymbol{\beta} = \boldsymbol{\beta}_{p_{1}} + \boldsymbol{\beta}_{p_{2}} + \ldots \boldsymbol{\beta}_{p_{m}} \mod q$.
Note that each $\boldsymbol{\beta}_{p_{i}}$ is a vector itself, containing a share of each of the coefficients $\beta_0, \beta_1, \ldots, \beta_k$.
\end{enumerate}

In step \ref{step4} and \ref{step5} above, all the $m$ parties perform computations and exchange encrypted messages with each other. To facilitate the communication among all parties, and to limit the number of communication channels (sockets) that need to be opened during execution, we use a Broadcast Agent (BA), more details about which are provided in Section \ref{sec:architecture}.

The five steps outlined above allow $m$ source parties to work together to train an LR model on all of their data. None of the source parties sees data from any of the other source parties in an unencrypted way, and none of the source parties can reconstruct the LR model by itself. Instead, at the end of the training protocol, shares of the coefficient vector are held in a distributed fashion by all $m$ parties. This entails that, when making new predictions with the trained model, all $m$ parties have to be involved, as we describe in Section \ref{sec:inti}.

\subsection{Inference for target-independent LR}\label{sec:inti}

During the \textit{inference phase} (Fig.~\ref{figure:SCEN1TEST}), the target party obtains a prediction for its input data by sending shares of its input to all the $m$ parties. The parties engage in an SMC protocol among themselves. At the end of the evaluation protocol, each of the $m$ parties sends its share of the result back to the target party, which adds the shares up to obtain the prediction. 

Concretely, inference in the target-independent LR scenario consists of the following four steps:
\begin{enumerate}[leftmargin=*,noitemsep,topsep=3pt]

\item Offline phase: The TI distributes the necessary correlated random numbers to the parties. These random numbers will be needed for secure multiplications in the third step.

\item Distribution of $\shareq{{\bf x}}$:  The target party maps the numbers in its input vector ${\bf x}$ to elements of the finite field $\mathbb{F}_q$ using the method described in \secref{sec:prelim}, and sends a share  ${\bf x}_{p_{i}}$ of the resulting vector ${\bf x}$ to each of the source parties, with
${\bf x} = {\bf x}_{p_{1}} + {\bf x}_{p_{2}} + \ldots +{\bf x}_{p_{m}} \mod q$.

\item Joint computation of $\shareq{\boldsymbol{\beta} \cdot {\bf x}^T}$: the $m$ source parties perform joint computations over their shares 
$\shareq{\boldsymbol{\beta}}$ from step \ref{step5} in Section \ref{sec:trti} and their shares $\shareq{{\bf x}}$ from step 2 above to obtain shares of 
$\shareq{\boldsymbol{\beta} \cdot {\bf x}^T}$. To this end, they use the secure matrix multiplication $\pmmul$ and secure truncation $\ptrunc$ protocols mentioned in \secref{sec:prelim}. Each source party sends it computed share, which we refer to as $\hat{y}_{p_i}$ below, back to the target party.

\item Local computation of $\hat{y}$: the target party adds the received shares $\hat{y}_{p_i}$ ($i = 1, \ldots, m$) to learn the prediction $\hat{y} = \hat{y}_{p_1} + \hat{y}_{p_2} + \ldots + \hat{y}_{p_m} \mod q$.
\end{enumerate}

\subsection{Training for target-calibrated LR}\label{sec:trtc}
PP target-calibrated LR requires $m$ cases of secure two-party computation during the \textit{training phase} (Fig.~\ref{figure:SCEN2TRAIN}), since each of the $m$ LR models is trained with data from only two parties, namely a source party and the target party. That means that instead of just one matrix $X$ as in Section \ref{sec:trti}, $m$ such matrices are implicitly used, which we denote as $X^{(1)}, X^{(2)}, \ldots, X^{(m)}$. Each such matrix $X^{(i)}$ consists of rows with training examples from source party $p_i$ and rows with calibration data from the target party. In practice, none of these matrices exist in one place. Instead, each matrix $X^{(i)}$ exists in a distributed fashion across source party $p_i$ and the target party, who each have a share of it. In addition, for each matrix $X^{(i)}$ there is a corresponding response value vector ${\bf y}^{(i)}$ which is shared in a similar way between source party $p_i$ and the target party. 

At the end of the training protocol, the coefficients of the $i$th regression model ($i = 1, \ldots, m$) are shared between the target party and source party $p_i$. These coefficients are computed through the steps described below.

\begin{enumerate}[leftmargin=*,noitemsep,topsep=3pt]
\item \label{step1c} Offline phase: The TI distributes the necessary correlated random numbers to the parties. These random numbers will be needed for secure multiplications in step \ref{step5c} and \ref{step6c} below.

\item \label{step2c} Local construction of $\shareq{X^{(i)}}$ and $\shareq{\vecy^{(i)}}$, $i=1,\ldots,m$: the target party maps its calibration data to elements of a finite field $\mathbb{F}_q$ using the method described in \secref{sec:prelim}, and creates an $n_{t} \times k$ matrix $X_{t}$ and a vector ${\bf y}_{t}$ of length $n_t$, with $n_t$ the number of training examples in the calibration data. 
Likewise, each source party $p_i$ maps its fixed-point inputs to elements of $\mathbb{F}_q$, and creates an $n_{p_i} \times k$ matrix $X_{p_i}$ and a vector ${\bf y}_{p_i}$ of length $n_{p_i}$, with $n_{p_i}$ the number of training examples owned by party $p_i$ and $k$ the number of features. 
At this point, the source parties and the target party are holding shares of the matrices $X^{(i)}$ and vectors ${\bf y}^{(i)}$ for $i= 1, \ldots, m$ in a distributed fashion such that
$X^{(i)}  = X_{p_i} + X_{t} \mod q \mbox{\ and\ } 
{\bf y}^{(i)}  = {\bf y}_{p_i} + {\bf y}_{t} \mod q.$ Note that these matrices are never constructed in their entirety in practice.

\item \label{step4c} Local computation of $\shareq{{(X^{(i)})}^T X^{(i)}}$ and $\shareq{{(X^{(i)})}^T\vecy^{(i)}}$, $i=1,\ldots,m$: 
each source party \textbf{locally} computes the $k \times k$ matrix $(X_{p_i})^T (X_{p_i})$ and the $k \times 1$ matrix $(X_{p_i})^T (\vecy_{p_i})$. The target party \textbf{locally} computes the $k \times k$ matrix $(X_t)^T (X_t)$ and the $k \times 1$ matrix $X_t^T \vecy_t$.

\item \label{step5c} Joint computation of $\shareq{({(X^{(i)})}^T X^{(i)})^{-1}}$, $i=1,\ldots,m$:
the target party separately engages in \textbf{joint} computations with each source party $p_i$ over their shares $\shareq{{(X^{(i)})}^T X^{(i)}}$ from step \ref{step4c} to compute shares $\shareq{({(X^{(i)})}^T X^{(i)})^{-1}}$ using the protocol for covariance matrix inversion $\pmatinv$ mentioned in Section \ref{sec:prelim}.

\item \label{step6c} Joint computation of $\shareq{\boldsymbol{\beta}^{(i)}}$, $i=1,\ldots,m$:
the target party separately engages in  
\textbf{joint} computations with each source party $p_i$ over their shares $\shareq{({(X^{(i)})}^T {X^{(i)}})^{-1}}$ from step \ref{step5c} and shares $\shareq{{(X^{(i)})}^T\vecy^{(i)}}$ from step \ref{step4c} to compute 
shares $\shareq{\boldsymbol{\beta}^{(i)}}$ of the coefficient vector $\boldsymbol{\beta}^{(i)}=({(X^{(i)})}^T {X^{(i)}})^{-1}{(X^{(i)})}^T\vecy^{(i)}$. 
To this end, they perform distributed matrix multiplication between $({(X^{(i)})}^T {X^{(i)}})^{-1}$ and ${(X^{(i)})}^T\vecy^{(i)}$ by performing secure matrix multiplication $\pmmul$ and secure truncation $\ptrunc$ operations.

\begin{figure}\begin{tcolorbox}[
    standard jigsaw,
    opacityback=0,  
]
\begin{center}
\includegraphics[width=3in]{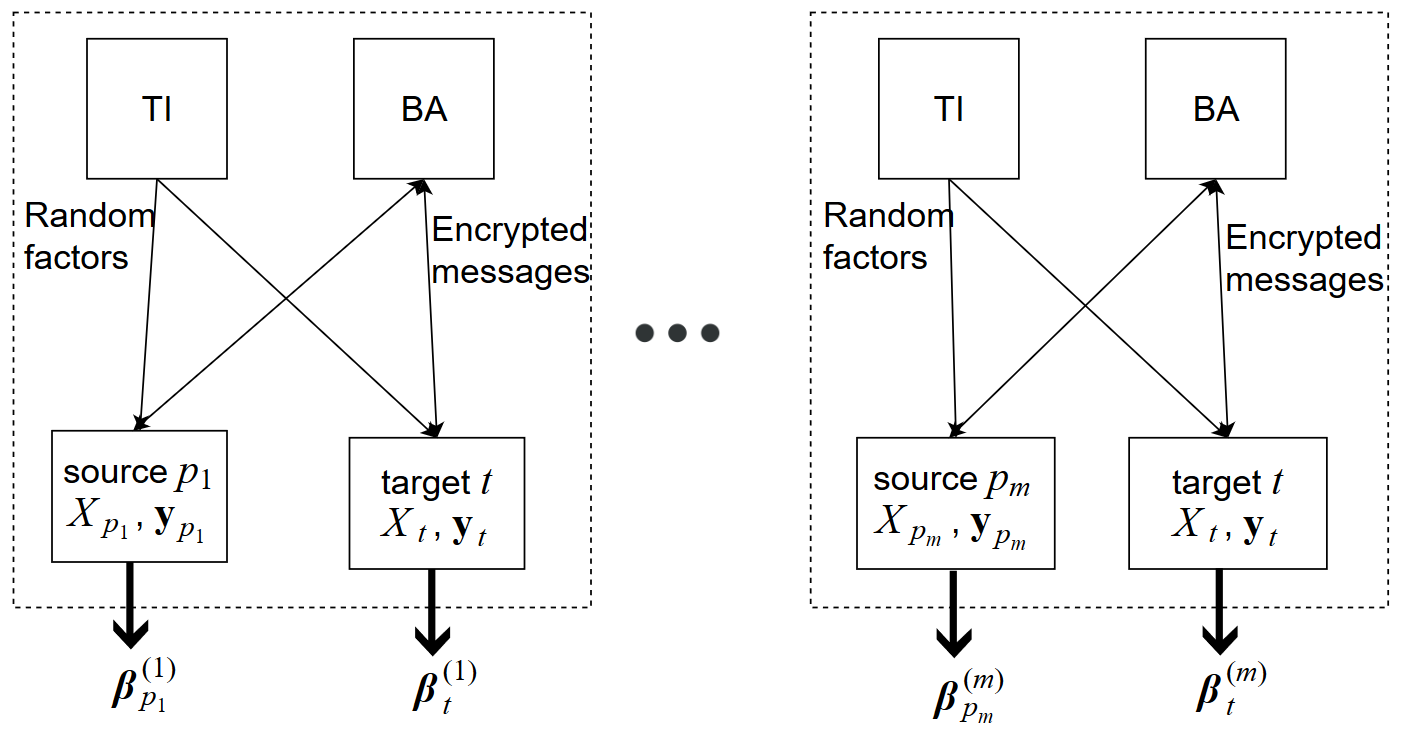}
\end{center}
\end{tcolorbox}
\caption{Training phase of privacy-preserving target-calibrated LR.}
\label{figure:SCEN2TRAIN}
\end{figure}

\begin{figure}\begin{tcolorbox}[
    standard jigsaw,
    opacityback=0,  
]
\begin{center}
\includegraphics[width=3in]{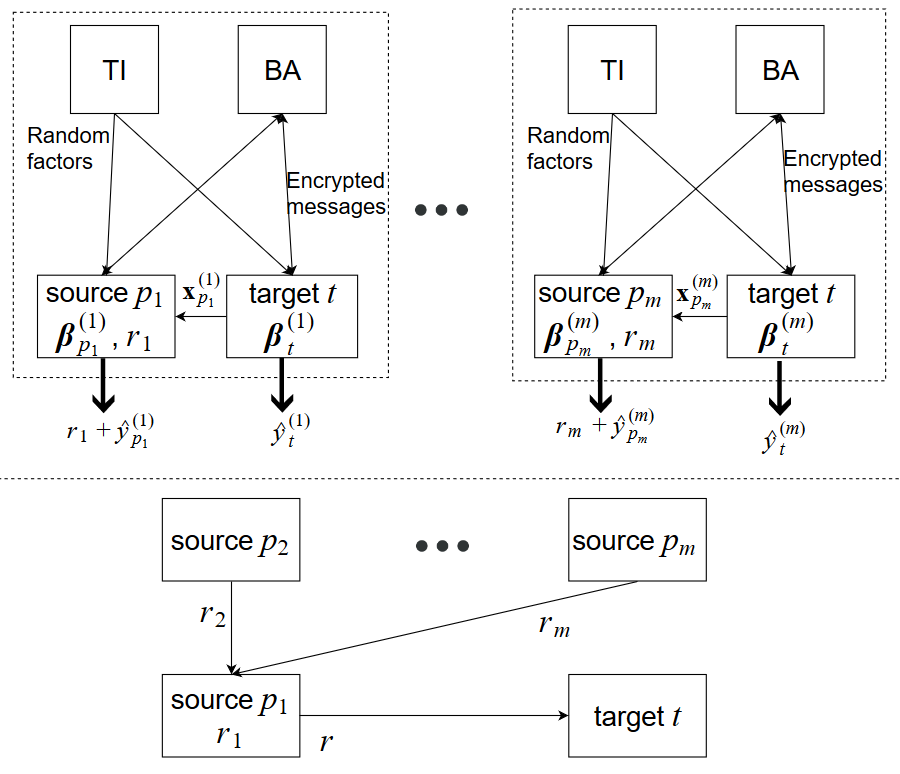}
\end{center}
\end{tcolorbox}
\caption{Inference phase of privacy-preserving target-calibrated LR.}
\label{figure:SCEN2TEST}
\end{figure}

When all computations are finished, $m$ LR models have been trained. The coefficient vector of the $i$th model ($i=1,\ldots,m$) is secret shared between the $i$th source party on one hand and the target party on the other hand, i.e., $
\boldsymbol{\beta}^{(i)} = \boldsymbol{\beta}^{(i)}_{p_i} + \boldsymbol{\beta}^{(i)}_{t} \mod q, \mbox{\ \ \ \ for\ } i = 1, \ldots, m$
\end{enumerate}

\subsection{Inference for target-calibrated LR}\label{sec:TCLRinference}

The \textit{inference phase} (Fig.~\ref{figure:SCEN2TEST}) for target-calibrated LR requires the computation of the average of the outputs of the $m$ regression models, which again involves secure two-party computations among the target party and each source party. As in Section \ref{sec:inti}, the target party has an input $\vecx$ for which it needs a prediction. To this end, the target party engages in a secure computation with each source party $p_i$ to construct a secret sharing $\shareq{\hat{y}^{(i)}}=\shareq{\boldsymbol{\beta}^{(i)} \cdot \vecx^T}$. The final prediction is the average of all the $\hat{y}^{(i)}$ values, $i=1,\ldots,m$. Instead of having each source party $p_i$ send its share of $\hat{y}^{(i)}$ to the target party, which would reveal information that is not strictly necessary, each of the parties $p_i$ first mask their prediction share by adding a random number $r_i$ and open the masked result to the target party. In addition, each party $p_i$ sends its random mask $r_i$ to one of the source parties ($p_1$ in Fig.~\ref{figure:SCEN2TEST}), which adds them up, and sends the result $r$ to the target party. Finally the target party locally adds up the masked shares of the $\hat{y}^{(i)}$ values (received directly from each of the source parties), subtracts the sum of the masks $r$ (received from the designated source party that is responsible for constructing this sum), adds its own shares of the $\hat{y}^{(i)}$ values, and divides by the number of source parties to obtain the final prediction. Concretely, inference in the target-calibrated LR scenario consists of the steps described below.
\begin{enumerate}[leftmargin=*,noitemsep,topsep=3pt]

\item Offline phase: The TI distributes the necessary correlated random numbers to the parties. These random numbers will be needed for secure multiplications in step 3.

\item Distribution of $\shareq{\bf x}$: the target party maps the numbers in its input vector to elements of the finite field $\mathbb{F}_q$ using the method described in \secref{sec:prelim}. Next the target party secret shares the input vector ${\bf x}$ with each of the source parties, possibly in a different way, i.e.
${\bf x}^{(i)} = {\bf x}^{(i)}_{p_i} + {\bf x}^{(i)}_{t} \mod q$ for $i = 1, \ldots, m$.

\item \label{step3d} Joint computation of $\shareq{\hat{y}^{(i)}}$, $i=1,\ldots,m$: the target party performs joint computations with each source party $p_i$ over their shares
$\shareq{\boldsymbol{\beta}^{(i)}}$ from step \ref{step6c} in Section \ref{sec:trtc} and their shares $\shareq{\bf x}$ from step 2 above to compute shares  
$\shareq{\hat{y}^{(i)}} = \shareq{\boldsymbol{\beta^{(i)}} \cdot {\bf x}^T}$. To this end, the parties use the secure matrix multiplication $\pmmul$ and secure truncation $\ptrunc$ protocols mentioned in \secref{sec:prelim}. At the end of this step, shares of the predictions made by each of the $m$ LR models have been constructed and are held, in a distributed fashion, by the target party and each of the source parties: $\hat{y}^{(i)} = \hat{y}^{(i)}_{p_i} + \hat{y}^{(i)}_{t} \mod q$ for $i = 1, \ldots, m$.

\item \label{step4d} Local masking of shares: all the source parties mask their share $\hat{y}^{(i)}_{p_i}$ by adding a random value $r_i$, and subsequently send the masked prediction to the target party $t$.

\item \label{step5d} Local computation of sum of masks: all the source parties send their random masks $r_i$, to one of the parties (chosen based on the asymmetric bit as mentioned in Section \ref{sec:architecture}). This designated party adds up all the random masks $r_i$, $i=1,\ldots,m$, and sends the result to the target party. 

\item \label{step6d} Local computation of final prediction:
the target party adds up all the masked prediction shares received in step \ref{step4d}, i.e.~$\hat{y}^{(i)}_{p_i}$ + $r_i$, $i=1,\ldots,m$, subtracts the sum of the random masks received in step \ref{step5d}, adds its own shares $\hat{y}^{(i)}_{t}$, $i=1,\ldots,m$ computed in step \ref{step3d}, and takes the average.
\end{enumerate}

%
%
\section{Implementation in Lynx}\label{sec:architecture}
In this section we describe design decisions made when implementing the protocols from Section \ref{sec:protocols} in Lynx \cite{Lynx},
a framework that we developed for SMC.
As explained in Section \ref{sec:prelim}, our protocols are developed for the commodity-based model, where the players running the distributed computations receive pre-distributed data from a trusted source (TI) during a setup phase. This data consist of correlated random numbers that help to mask information during the computations. The algorithms for secure LR described in Section \ref{sec:protocols} rely on the cryptographic protocols for secure matrix multiplication, matrix inversion, and truncation that were mentioned in Section \ref{sec:prelim}. We implemented these protocols such that the performance scales with an increasing number of players involved in the computations. The ability to efficiently accomodate more than three parties to jointly perform the computations, sets our Lynx framework apart from existing SMC frameworks that are limited to two or three parties \cite{Bogdanov:2008,2018arXiv180103239S} or that become computationally heavy with more than three parties \cite{Ben-David:2008}.

\begin{figure}\begin{tcolorbox}[
    standard jigsaw,
    opacityback=0,  
]
\centering
\includegraphics[width=2.4in,height=2in]{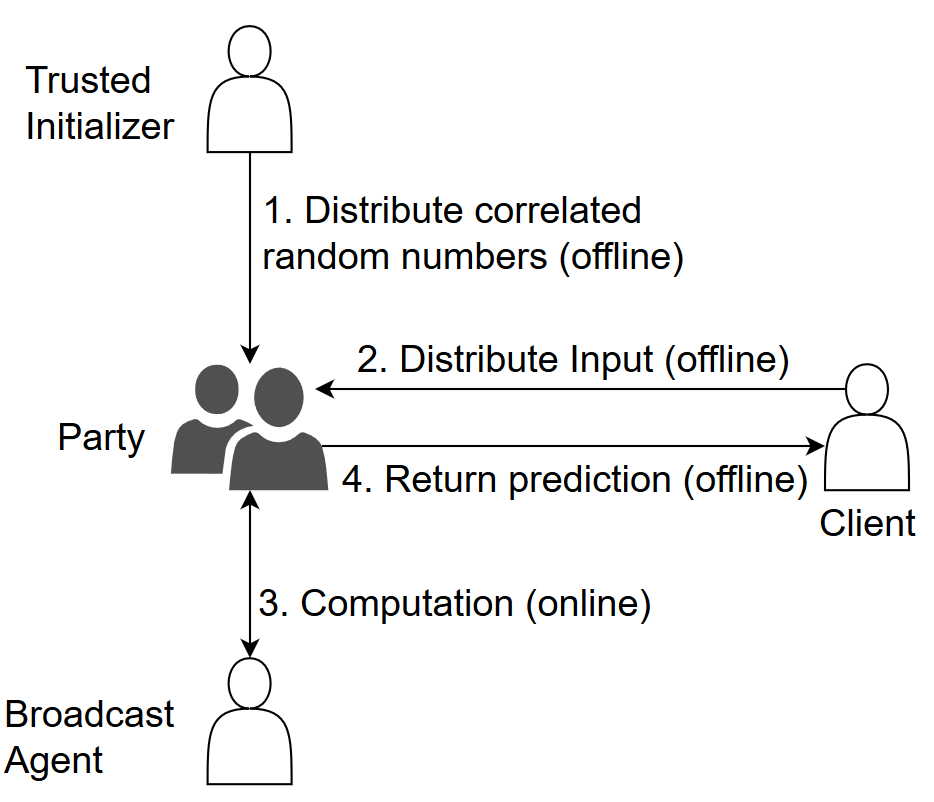}
\end{tcolorbox}
\caption{Different roles in the SMC framework Lynx.}
\label{figure:actor_roles}
\end{figure}

There are four significant roles that run at various stages for end-to-end model training and inference. They function as illustrated in Fig.~\ref{figure:actor_roles}. A deployed system consists of two or more Parties, one Broadcast Agent, one Trusted Initializer, and one or more Clients.  The Parties communicate via the Broadcast Agent. The difference between a Party and a Client is that a Party engages in SMC computations, while a Client does not. The role of the latter is limited to distributing input data and receiving corresponding outputs that were computed by the Parties in a secure way. In the target-independent LR scenario for driver drowsiness prediction (see Section \ref{sec:results}) for instance, the target driver is a Client, while in the target-calibrated LR scenario, the target driver is a Party as well as a Client.

\subsubsection{Party}
The Party is the core module responsible for model training and inference. The Parties take shares of the input, compute the result of a function over this input, and return the output shares. If Lynx is used for training, they produce shares of the trained model as an output, such as the shares of the coefficient vector of a LR model. When used for prediction, they produce the shares of the predicted result. At no point does any of the individual Parties know the data or the result held by any other Party in an unencrypted way.

\subsubsection{Trusted Initializer}

The TI runs as an offline program to generate the set of correlated random data required for the computations. It passes shares of this data to all the parties before the start of the computations and does not interfere with the computation any further from that point onward. 

\subsubsection{Broadcast Agent}

One of the cornerstones of SMC is the pair-wise exchange of masked data between the Parties involved in the computations. While this works well in a 2-party scenario, the performance can get worse with an increase in the number of Parties, which is a plausible explanation for why most existing SMC implementations only have few parties. In Lynx we have introduced a ``bulletin board'' functionality referred to as the Broadcast Agent. It is a dummy server which relays public messages to all Parties. The principal benefit of using the Broadcast Agent is to reduce the number of communication channels (sockets) that need to be opened, thereby greatly enhancing the efficiency of the communication. A traditional broadcast protocol would establish $O(m^2)$ sockets among the $m$ parties, while only $O(m)$ sockets are necessary when a Broadcast Agent is used.

\subsubsection{Client}

The Client is the user that holds the private data and wants to get predictions from the model. The client secret shares the input among $m$ Parties, such that none of the Parties knows the actual input, and receives back the shares of the predicted value. This way neither Party gets to know anything about the client's input or prediction.
In the case of a target-calibrated LR, the target driver acts like a Party while training the model, and acts like a Client during the inference. 

\setcounter{subsubsection}{0}

The Broadcast Agent and Trusted Initializer may exist on one or more servers. The Parties can run on a single or distributed network.
Lynx uses two main architecture patterns: 1) Client-server architecture for all communications of the parties with the trusted initializer and the Broadcast Agent; 2) Microservices architecture to achieve modularity between all the SMC protocols. This allows to reuse the protocols and run them concurrently at different stages of computations. The Parties jointly compute an ML model (LR in this paper) by calling the different cryptographic protocols as microservices. Lynx is designed such that independent computations can happen in parallel, thus increasing throughput. Finally, we have created a number of utility protocols in Lynx which help in batch processing many cryptographic protocols to reduce communication overhead among parties.

%
%

\section{Experimental Results}\label{sec:results}

\subsection{Dataset and Hardware Specifications}\label{sec:dataset}
We evaluated the implementation of our cryptographic protocols using the same data and scenarios for detecting driver drowsiness based on EEG signals as Wu et al.~\cite{wu2017driver}. We used data from subjects
who participated in a 60-90 minutes sustained-attention driving experiment in a real vehicle mounted on a motion platform immersed in a 360-degree virtual-reality scene. To induce drowsiness during driving, the
virtual-reality scenes simulated monotonous driving at a fixed
100 km/h speed on a straight and empty highway. During
the experiment, lane-departure events were randomly applied
every 5-10 seconds, and participants were instructed to steer
the vehicle to compensate for these perturbations as quickly
as possible. 16 voluntary participants of age 24.2 $\pm$ 3.7 (10 males and 6 females) with normal or corrected-to-normal vision were recruited in this study. Data from one subject was not correctly recorded, so we used only 15 subjects. 

We defined a function \cite{wei2015selective} to map the recorded response time $\tau$ to a drowsiness index $y\in[0, 1]$:
\begin{align}
y=\max\left\{0,\,(1-e^{-(\tau-\tau_0)})/(1+e^{-(\tau-\tau_0)})\right\}
\end{align}
$\tau_0=1$ was used in this paper, as in \cite{wei2015selective,wu2017driver}. The drowsiness indices were then smoothed using a 90-second square moving-average window to reduce variations. This does not reduce the sensitivity of the drowsiness index because the cycle lengths of drowsiness fluctuations are longer than 4 minutes \cite{makeig1993lapse}. 

During the experiment, the participants’ scalp EEG signals were recorded using a 32-channel (30-channel EEGs plus 2-channel earlobes) 500 Hz Neuroscan NuAmps
Express system (Compumedics Ltd., VIC, Australia). Afterwards, the EEG data was preprocessed and features were extracted, resulting in a sequence of 1200 epochs for each driver, in which each epoch is characterized by 30 numerical values extracted from the EEG signal. For each of the 15 drivers we therefore have a dataset consisting of 1200 rows, in chronological order, each consisting of 30 numerical input values and a response value (the level of drowsiness). 
For more details on the preprocessing of the data, we refer to Wu et al.~\cite{wu2017driver}. 

The experiments documented below were run on a AWS c5.9xlarge machines with 36 vCPUs, 72.0 GiB Memory. Each of the Trusted Initializer, Broadcast Agent, and all Parties ran on separate machines. Each runtime experiment was repeated 3 times and average results are reported.

\subsection{Results for Target-Independent LR}
We train an LR model with data from $m$ source drivers (Fig.~\ref{figure:SCEN1TRAIN}) and apply it to make inferences about a new target subject (Fig.~\ref{figure:SCEN1TEST}). Since each source driver has 1200 rows of data, the full matrix $X$ from Equation (\ref{EMRformula}) is a $(m \cdot 1200) \times 30$ matrix, while ${\bf y}$ is a $(m \cdot 1200) \times 1$ vector. At no point $X$ and $\vecy$ are constructed in full. Each source party naturally has a share of $X$ and ${\bf y}$ at the outset of the algorithm: a $1200 \times 30$ matrix $X_{p_i}$ and a $1200 \times 1$ vector ${\bf y}_{p_i}$ with the data from driver $i$. 

The first columns of Table \ref{tab:resultsTILR} contain runtime results for training a target-independent LR model with data from $m$ source drivers, as the number of source drivers increases from $m=2$ up to $m=14$. In the clear, i.e.~without any encryption, training is very fast and completes within a fraction of a second. As expected, training in a PP fashion using SMC is computationally heavier. The runtime grows with the number of drivers, because there is more training data available that needs to be processed, and more parties that need to communicate and coordinate. Still, as is clear from Table \ref{tab:resultsTILR}, an increase in the number of parties has a moderate impact on the runtime, demonstrating that the implementation in Lynx of the PP protocol for training a LR regression model is scalable.

\begin{table}
\caption{Results for training and inference in the target-independent LR scenario with an increasing number of parties (drivers).}\label{tab:resultsTILR}
\begin{center}
\begin{tabular}{|c|r|r|r|r|r|}
\hline
& \multicolumn{2}{c|}{\bf Training} & \multicolumn{3}{c|}{\bf Inference}\\
\hline
& \multicolumn{2}{c|}{\bf Runtime (sec)} & \multicolumn{2}{c|}{\bf Runtime (sec)} & {\bf RMSE}\\
\hline
\# of parties & In the clear & SMC & In the clear & SMC & \\
\hline
2 & 0.10 & 48.51 & 0.004 & 2.82& 0.051\\
3 & 0.15 & 77.55 & 0.004 & 3.25&  0.050\\
4 & 0.22 & 106.91 & 0.004 & 3.81& 0.043\\
5 & 0.28 & 132.24 & 0.004 & 4.43&  0.087\\
6 & 0.35 & 153.90 & 0.004 & 4.98& 0.129\\
7 & 0.42 & 171.87 & 0.004 & 5.73&  0.106\\
8 & 0.46 & 201.26 & 0.004 & 6.46& 0.090\\
9 & 0.49 & 225.58 & 0.004 & 7.03&  0.082\\
10 & 0.51 & 245.74 & 0.004 & 7.91& 0.074\\
11 & 0.52 & 280.96 & 0.004 & 8.42& 0.071\\
12 & 0.53 & 299.11 & 0.004 & 9.12& 0.071\\
13 & 0.45 & 328.67 & 0.004 & 10.03& 0.055\\
14 & 0.43 & 350.39 & 0.004 & 10.08& 0.048\\
\hline
\end{tabular}
\end{center}
\end{table}

Next we evaluate the predictive accuracy of the trained target-independent LR models. To this end, we treat driver 15, which was not used for training the models in Table \ref{tab:resultsTILR} as the target driver. We use the trained models to predict the response value for each of the 1200 rows in the data of the target driver. In the target-independent scenario, the coefficient vector $\boldsymbol{\beta}$ of the trained LR model is kept in a distributed fashion with each of the $m$ source parties involved in the training. Making PP predictions with the trained model is therefore an $m$-party SMC problem, the runtime of which grows with $m$, as shown in the ``Inference'' columns in Table \ref{tab:resultsTILR}.
The RMSE (Root Mean Square Error) for those predictions is reported in the last column of Table \ref{tab:resultsTILR}. We obtained the same RMSE in the clear as when computing over encrypted data, highlighting that there is no accuracy loss when computing in a PP way.

\subsection{Results for Target-Calibrated LR}
In the target-calibrated LR scenario, $m$ LR models are trained (Fig.~\ref{figure:SCEN2TRAIN}). For each LR model, the matrix $X^{(i)}$ consists of the 1200 rows from the $i$th source driver, followed by the first 100 rows of the target driver, which we use as calibration data. This means that each $X^{(i)}$ is a $1300 \times 30$ matrix, for $i = 1, \ldots, m$. Similarly, each ${\bf y^{(i)}}$ is a $1300 \times 1$ vector.

\begin{table}
\caption{Results for the target-calibrated LR scenario. The $i$th row in the table contains the results about the LR model trained with 1200 rows of data from the $i$-th driver ($i=1,\ldots,14$) combined with 100 rows of calibration data from driver 15.}\label{tab:resultsTCLR}
\begin{center}
\begin{tabular}{|c|r|r|r|r|r|}
\hline
& \multicolumn{2}{c|}{\bf Training} & \multicolumn{3}{c|}{\bf Inference}\\
\hline
& \multicolumn{2}{c|}{\bf Runtime (sec)} & \multicolumn{2}{c|}{\bf Runtime (sec)} & {\bf RMSE}\\
\hline
Source party id & In the clear & SMC & In the clear & SMC & \\
\hline
1 & 0.06& 51.23& 0.004& 2.61& 0.114\\
2 & 0.06 & 51.95& 0.003& 2.68& 0.045\\
3 & 0.06& 51.95&  0.003& 2.67& 0.145\\
4 & 0.06& 51.88& 0.004& 2.71& 0.055\\
5 & 0.06& 51.62& 0.003& 2.71& 0.086\\
6 & 0.06& 51.41& 0.003& 2.64& 0.097\\
7 & 0.06& 51.57& 0.004& 2.65& 0.062\\
8 & 0.06& 51.49& 0.003& 2.64& 0.045\\
9 & 0.07& 52.06& 0.004& 2.62& 0.066\\
10 & 0.06& 51.92& 0.003& 2.64& 0.078\\
11 & 0.06& 51.83& 0.003& 2.66& 0.057\\
12 & 0.06& 52.31& 0.003& 2.68& 0.194\\
13 & 0.06& 51.45& 0.003& 2.61& 0.186\\
14 & 0.06& 51.50& 0.003& 2.65& 0.053\\
\hline
All & 0.07 & 52.00& 0.008& 2.89& 0.048\\
\hline
\end{tabular}
\end{center}
\end{table}

Table \ref{tab:resultsTCLR} present the runtimes for training target-calibrated regression models with calibration data from a target driver (in this case, driver 15) combined with data from one of the source drivers. Since training each regression model only involves two parties (the target and one of the source drivers), this is a 2-party computation. As shown in Table \ref{tab:resultsTILR}, the average runtime for training a regression model with two parties is around 51.73 sec. As all 14 models can be trained in parallel, the training time to learn the entire target-calibrated model is approximately 52 sec. We evaluate the predictive accuracy of the trained target-calibrated LR models when predicting the response value for each of the remaining 1100 rows in the data of the target driver, i.e.~the rows that were not used as calibration data. The RMSE for those predictions is reported in the last column of Table \ref{tab:resultsTCLR}, along with the time needed to make those predictions. The final prediction is computed as the average of the predictions of all $m=14$ LR models.
In the SMC based approach, an additional time of 0.24 sec is required for all parties to mask their prediction shares, to send the masked prediction shares to the target party, to send the mask to one of the parties, and to allow the target party to compute the final result (cfr.~step \ref{step4d} to \ref{step6d} in Section \ref{sec:TCLRinference}). The time  on average to make a prediction for the 1100 rows of a target driver is 0.008 sec when done in the clear, i.e.~without encryption, and $2.65 + 0.24$ i.e 2.89 sec when done in a PP way using SMC. The RMSE is the same whether the predictions are made with full exposure of the EEG data or in private.

\section{Conclusion}
This work presented the first application of commodity-based SMC for privacy-preserving processing of EEG data, as well as the largest documented experiment of secret sharing based SMC in general, with 15 players involved in all the computations. We proposed algorithms for PPLR in a target-independent as well as a target-calibrated scenario. We have implemented these algorithms in Lynx, a new SMC framework that we created to enable efficient SMC among many parties. The runtime results of our experiments for predicting driver drowsiness show that our LR protocols and their implementation scale very nicely with an increasing number of drivers involved in the computations, and that the privately trained LR models are as accurate as those trained in the clear, i.e.~without any encryption. 
Our work shows that additive secret sharing based SMC is a viable mechanism for protecting the privacy of users in future brain-computer interface applications. However, our running times were obtained using powerful machines and much work is needed to make these protocols practical in constrained computing devices. Interesting future research directions include: (i) to design protocols that work for more restrictive adversarial models (such as fully malicious or covert) and (ii) to improve communication, computational and round complexities for our protocols.

\end{document}